\documentclass[a4paper,11pt]{article}

\usepackage{bm}
\usepackage{physics}
\usepackage{amsmath, amssymb}  % For mathematical symbols and equations
\usepackage{graphicx}          % For including graphics
\usepackage{natbib}[round]

\usepackage[margin=18truemm]{geometry}

\usepackage{float}             % For controlling float placement
\usepackage{tcolorbox}         % For colored boxes
\usepackage{multicol}          % For handling multiple columns
\usepackage{diagbox}

\title{Effect of magnetic drift on the stability structure of the ambipolar condition}
\author{Keiji Fujita and Shinsuke Satake \\
        \small National Institute for Fusion Science (NIFS) \\
        \small \texttt{fujita.keiji@nifs.ac.jp}}
\date{\today}

\usepackage{hyperref}
\hypersetup{%
 setpagesize=false,%
 bookmarksnumbered=true,%
 colorlinks=true,%
 linkcolor=blue,%
 citecolor=blue,%
 pdftitle={},%
 pdfsubject={},%
 pdfauthor={},%
 pdfkeywords={}}

\begin{document}

\maketitle

\begin{abstract}
    In non-axisymmetric plasmas, the ambipolar condition may have multiple roots.
    In such cases, the evolution of the ambipolar electric field can be described by the dynamics in a bistable potential, where the relative depth of the potential wells primarily determines the realized root. 
    In this study, we show that the inclusion of the magnetic drift in the orbit model can significantly modify the potential landscape and affect root selection.
    This effect provides a possible explanation for discrepancies between simulation results obtained using different orbit models, as well as between simulations and experimental observations of ambipolar radial electric field profiles.
    Further, the analysis suggests that the ambipolar electric field may be more susceptible to fluctuations than previously expected, indicating the potential relevance of noise-induced state transitions.
\end{abstract}

\section{Introduction}

The ambipolar condition in a non-axisymmetric fusion plasma may have multiple roots. 
In such cases, the evolution of the ambipolar electric field can be described as the dynamics of an overdamped particle in a bistable potential.
The plasma is stable when the radial electric field is located at one of the potential minima, which correspond to solutions of the ambipolar condition.
When the effects of noise are taken into account, the notion of stability is replaced by the probability of finding the system in a given state and by the transition probabilities between stable states.
Problems with this structure have been studied in many areas of science, particularly chemical and biochemical contexts \citep{hanggi1990reaction}.

\cite{shaing1984stability,shaing1984noise} revealed this structure and formulated the theoretical framework for analyzing the ambipolar electric field dynamics.
However, in our view, the implications of these studies have not been fully appreciated.
In this study, we therefore revisit these works and show why the non-axisymmetric neoclassical transport system should be analyzed from the perspectives of ambipolar electric field dynamics.
We then discuss and demonstrate how the inclusion of the magnetic drift in drift-kinetic orbit models modifies the macroscopic transport by altering the root selection of the ambipolar condition.
This analysis provides a possible explanation for discrepancies in ambipolar electric field profiles obtained using different drift-kinetic simulation models \citep{fujita2020global}, as well as for differences between simulations and experimental observations \citep{ido2010experimental}.
The result also suggests that the radial electric field is much more susceptible to fluctuations than previously expected.
Furthermore, based on this insight, we briefly discuss the possibility of artificial control of the ambipolar electric field by means of noise-induced transitions.

The main subject of this study is a potential function that governs the stability of steady states of nonaxisymmetric neoclassical systems.
To clarify the physical meaning and significance of the potential function, we begin with a brief review of the stability analysis of thermodynamic systems in Sec~\ref{sec:thermodynamics}.
In the context, the connections between the ambipolar condition, some thermodynamic principles regarding entropy production, and the potential function are discussed.
Then, in Sec.~\ref{sec:ambipolar}, we revisit and summarize the works of \cite{shaing1984stability,shaing1984noise} based on the thermodynamic argument.
At the end of the section, we explain why the inclusion of the magnetic drift is crucial for analyzing neoclassical systems from the viewpoint of the stability structure. 
In Sec.~\ref{sec:simulation}, we demonstrate this point by presenting an example in which the most stable root changes from an ion-root to an electron-root when the magnetic drift effect is included.
Finally, the results are summarized and their implications are discussed in Sec.~\ref{sec:summary}.

%=============
\section{Stability of thermodynamic systems\label{sec:thermodynamics}}
%=============

\subsection{Prigogine's theorem and universal criterion of evolution}

Let us begin with a brief review of the basic argument concerning steady states and their stability of thermodynamic systems. 
In classical thermodynamics, steady states and their stability are determined by the extremal properties of the thermodynamic potentials, such as entropy and free energy.
In terms of entropy, a thermodynamic equilibrium state is characterized by $dS=0$, and its stability by the condition $\delta^2 S<0$ [see e.g. \citep{landau1980statistical,callen1991thermodynamics}].
For nonequilibrium systems, such functions cannot be found in general. 

In exploring the possibility of extending the principle to nonequilibrium systems, \cite{glansdorff1954proprietes} derived the following properties of entropy production.
Thermodynamic entropy production of a nonequilibrium steady system is given by
\begin{equation}
    \sigma
    =
    -\sum_{k} J_k X_k,
\end{equation}
where $J_k$ and $X_k$ are $k$-th macroscopic flux and conjugate driving force, respectively.
For a neoclassical transport system, the fluxes $J_k$ typically correspond to the particle flux $\Gamma_a$ and the energy flux (divided by the temperature $T_a$) $Q_a/T_a$.
The conjugate forces are then $\mu_a'/T_a-Z_aeE_r/T_a$ and $T_a'/T_a$, respectively, where $Z_ae$ is the charge, $E_r$ is the radial electric field, $\mu_a=T_a\ln n_{a0}-3T_a\ln T_a/2$ is the chemical potential with the background density $n_{a0}$, and a prime denotes the radial derivative.
The subscript $a$ refers to the species.

Let us decompose the variation of the thermodynamic entropy production as
\begin{equation}
    d\sigma
    =
    d_X\sigma
    +d_J\sigma,
\end{equation}
with
\begin{equation}
    d_X\sigma 
    =
    -\sum_k 
    J_k dX_k,
    \quad
    d_J\sigma
    =
    -\sum_k 
    X_k dJ_k.
\end{equation}
Correspondingly, the change of the total entropy production in the total volume $V$ of the system, $P=\int_V dV \sigma$ is split as $dP=d_XP+d_JP$.
Then, it can be shown that
\begin{equation}
    d_XP
    =
    -\int_V dV \sum_k J_k dX_k
    \leq 0,
\end{equation}
holds during the evolution of the system.
This is called the \textit{universal criterion of evolution}.
On the other hand, no general statement can be made for the evolution of the other part, 
\begin{equation}
    d_JP
    \equiv 
    -\int_V dV \sum_k X_k dJ_k.
\end{equation}
Thus, the total entropy production $P$ does not necessarily evolve toward its minimum value.

For a linear system, in which the transport coefficients defined by the phenomenological relation, $J_k=-\sum_l L_{kl}X_l$, are independent of $\{X_k\}$ and satisfy the Onsager relation
\begin{equation}
\label{eq:onsager_relation}
    L_{kl}=L_{lk},
\end{equation}
one finds that $d_J\sigma=\sum_{kl}L_{kl}X_kdX_l=d_X\sigma$.
The universal criterion then reduces to Prigogine's theorem \citep{prigogine1945moderation}
\begin{equation}
\label{eq:prigogine_theorem}
    d\sigma=2d_X\sigma=2d_J\sigma\leq 0.
\end{equation}
That is, such a linear nonequilibrium system evolves toward the state with the minimum value of entropy production.
The theorem is thus called the \textit{principle of minimum entropy production} \citep{gyarmati1970non,katchalsky1965nonequilibrium} as well.

\subsection{Generalized entropy production}

The universal criterion of evolution is a general criterion to constrain the evolution of thermodynamic systems including nonlinear systems.
However, the function $d_XP$, or its integrand
\begin{equation}
    d\Psi 
    \equiv d_X \sigma
    =
    -\sum_k J_kdX_k,
\end{equation}
does not usually serve as a nonequilibrium counterpart to thermodynamic potentials, since $d\Psi$ is not a total differential in general: the condition for the differential to be exact is
\begin{equation}
\label{eq:total_differential_condition}
    \frac{\partial J_k}{\partial X_l}
    =
    \frac{\partial J_l}{\partial X_k},
\end{equation}
for all $k$ and $l$ \citep{schutz1980}, which is not generally satisfied, unless the transport coefficients are independent of $\{X_k\}$ and (\ref{eq:total_differential_condition}) reduces to the Onsager relation (\ref{eq:onsager_relation}).
Thus, $d_X\sigma$ or $d_XP$ cannot be considered as the variation of a potential function \citep{nicolis1970thermodynamic}.

Therefore, \cite{glansdorff1962variational} and \cite{glansdorff1964general} investigated the condition for $d\Psi$ to be a total differential. 
Through the investigation, the concept of \textit{generalized entropy production} was introduced as a quantity that locally plays the same role as the entropy production for linear systems \citep{glansdorff1964general,prigogine1967introduction}.
If $d\Psi$ is integrable, the integrated function can be identified as a generalized entropy production.
When it cannot be integrated directly, a suitable integral factor should be found, analogously to the Carathéodory's argument for the second law of thermodynamics \citep{katchalsky1965nonequilibrium,schutz1980}.

\subsection{Implications for neoclassical transport}

To apply Prigogine's theorem to a linear neoclassical system in general, a slight modification is required. 
This is because the neoclassical transport coefficients defined through the relation $\sigma_a=\sum_b\sum_{kl}L_{kl}^{ab}X_{ka}X_{lb}$ do not generally satisfy the Onsager relation when the species $a$ and $b$ have unequal temperatures.
Instead, $D_{kl}^{ab}\equiv L_{kl}^{ab}/T_b$ generally satisfy the relation \citep{sugama1996entropy}
\begin{equation}
\label{eq:neoclassical_onsager_relation}
    D_{kl}^{ab}
    =
    D_{lk}^{ba},
\end{equation}
In this case, the relevant quantity becomes the so-called \textit{dissipation function} \citep{katchalsky1965nonequilibrium},
\begin{equation}
    \sum_a T_a\sigma_a
    =\sum_{a,b}\sum_{k,l}
    T_a L_{kl}^{ab}X_{ka}X_{lb}
    =\sum_{a,b}\sum_{k,l}
    D_{kl}^{ab}A_{ka}A_{lb},
\end{equation}
where $A_{ka}\equiv X_{ka}T_a$, instead of the entropy production $\sum_a\sigma_a$.

Since on the neoclassical time scale, the ambipolar electric field is usually the only time-varying driving force, taking the derivative of this function with respect to $E_r$ gives
\begin{equation}
\label{eq:minEPP_ambipolarity}
    \frac{\partial}{\partial E_r}\sum_a T_a\sigma_a
    =
    \sum_a Z_a e\Gamma_a
    \equiv 
    J_r
    =
    0,
\end{equation}
where, $J_r$ is the radial current.
This relation holds regardless of the temperature differences.
Thus, the stable stationary state of a linear neoclassical system coincides with the state where the radial current $J_r$ vanishes, i.e., when the intrinsic ambipolarity is satisfied.

Similarly, if we consider the integral of the dissipation function
\begin{equation}
    \mathcal{P}
    \equiv
    \sum_a
    \int_V T_a\sigma_a dV,
\end{equation}
the correspondence to (\ref{eq:universal_criterion}) is straightforwardly obtained from the evolution equation of $E_r$ [which will be given by (\ref{eq:ambipolar_Er_evolution})] as
\begin{equation}
\label{eq:universal_criterion}
    \frac{d_X \mathcal{P}}{dt}
    \equiv 
    \int_V J_r\frac{\partial E_r}{\partial t}dV
    =
    -\int_V 
    \frac{1}{\varepsilon}|J_r|^2
    dV
    \leq 0.
\end{equation}
This result states that the system evolves toward a state with vanishing radial current $J_r$.
This is nothing but another expression of the ambipolar condition, being a natural extension of the minimum dissipation property proved above.

Note that $d\Psi(E_r)\equiv J_rdE_r$ is trivially integrable, since the problem is one-dimensional, and we can define a potential function,
\begin{equation}
\label{eq:generalized_potential_Er}
    \Psi(E_r)
    \equiv 
    \int^{E_r} J(E_r') dE_r',
\end{equation}
which can be regarded as the generalized entropy production of Glansdorff and Prigogine.
\cite{shaing1984stability} called the corresponding function \textit{generalized heat production rate}, since its dimension is the same as entropy production times temperature.
Thus, the steady state of the neoclassical system can be found by the variation of $\Psi$, and its stability is determined by the sign of $\delta^2\Psi$.

A reason we included this section in the article is that we have encountered interpretations of $\Psi$ that may obscure its physical significance. 
For example, while Ref. \citep{kuczynski2025assessment} provides a comprehensive assessment of the ambipolar electric field profiles, the extremal properties of $\Psi$ are discussed in connection with the minimum entropy production principle (\ref{eq:minEPP_ambipolarity}), and the discrepancies between predicted and simulated transition locations are attributed to limitations of that principle.
However, as clarified above, these properties are not approximate, but follow exactly from the generalized form of Prigogine's theorem, i.e., the universal criterion of evolution and integrability of $d\Psi$.
This point is crucial for the arguments in the following sections, where we will discuss the role of $\Psi$ in neoclassical transport.

It should also be noted that the Onsager symmetry and consistent definitions of neoclassical transport coefficients and driving forces \citep{sugama1996entropy} had not yet been established at the time of \cite{shaing1984stability}. 
This may partly explain why the exact applicable conditions of the thermodynamic principles regarding the entropy production, or the dissipation function, and their relations to the ambipolar condition, have remained somewhat ambiguous.
Additionally, the notation of equation (2) of \cite{shaing1984stability} may be confusing.
The derivative with respect to the radial electric field appearing in the middle expression in the equation should be understood as the variation with respect to the driving force, in the sense of $d_X$ used in the present article.
Through the arguments presented in this section, the physical principle underlying the generalized heat production can be more clearly understood.

\section{Ambipolar condition\label{sec:ambipolar}}

\subsection{Deterministic model}

The evolution of the radial electric field $E_r$ is described by
\begin{equation}
\label{eq:ambipolar_Er_evolution}
    \varepsilon\frac{\partial E_r}{\partial t}
    =
    -J_r,
\end{equation}
where $\varepsilon$ is the permittivity.
If $J_r$ depends on time only through $E_r$, the right-hand side can be expressed in terms of a potential function, $\Psi=\int^{E_r} J_r(E_r')dE_r'$, which coincides with the function defined by (\ref{eq:generalized_potential_Er}).
Then, the generalized heat production rate $\Psi$ now appears as a potential function for the overdamped dynamics,
\begin{equation}
\label{eq:ambipolar_Er_evolution_potential}
    \varepsilon\frac{\partial E_r}{\partial t}
    =
    -\frac{\partial\Psi}{\partial E_r}.
\end{equation}
Therefore, an ambipolar state $(\partial E_r/\partial t=0)$ corresponds to a stationary state of the entire neoclassical system.
In analogous dynamical systems, the corresponding function is called by various names depending on the context, such as kinetic potential \citep{hochberg2021entropic} and effective potential \citep{bialek2000stability,endres2017entropy}.
Here, we simply call it a potential (or potential function) to avoid confusions with similarly named concepts that appear in the present context or in our future publications.

\begin{comment}
If a diffusion term of the form $f_D\equiv \partial^n (D(E_r)E_r/\partial r$ is included, 
It is still integrable.
However, the potential function is modified to
\begin{equation}
    \Psi'(E_r)
    =
    \int^{E_r}
    [J_r(E_r')
    -f_D(E_r)]dE_r'
\end{equation}
then, it does not coincides with the generalized heat production rate. 
\end{comment}

\begin{figure}[h]
    \centering
    \includegraphics[width=0.6\textwidth]{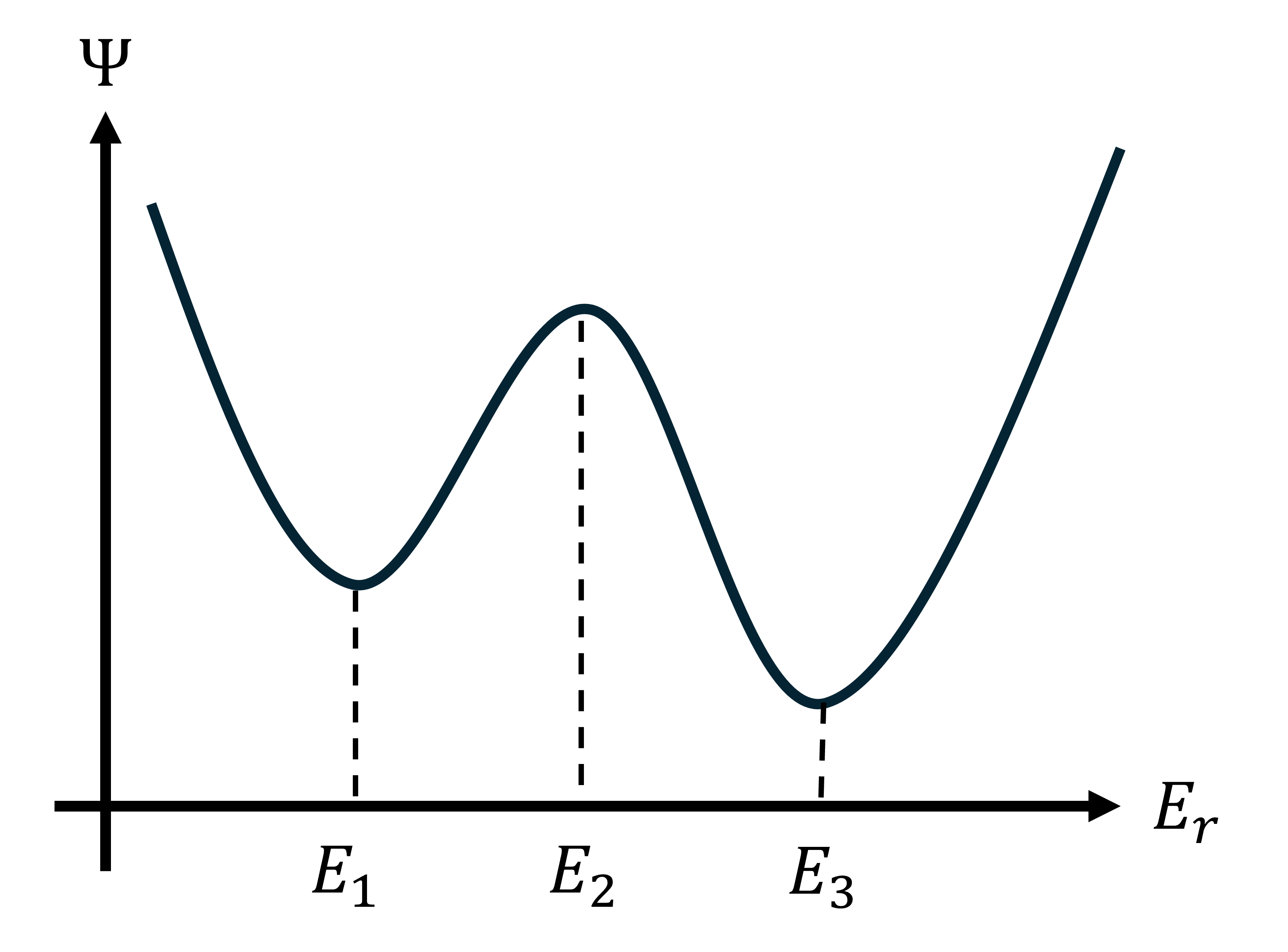}
    \caption{Schematic illustration of the potential function $\Psi(E_r)$ for a nonaxisymmetric plasma with two minima at $E_1$ and $E_3$, and a local maximum at $E_2$.}
    \label{fig:bistable_potential}
\end{figure}

There are at most three roots that satisfy the ambipolar condition $\partial E_r/\partial t=\partial\Psi/\partial E_r=0$.
For a root to be stable, $E_r$ must return to the root position when a small perturbation is applied, which requires $\partial^2\Psi/\partial E_r^2>0$ at the root.
This can be easily understood by interpreting the problem as that of an overdamped particle moving along the one-dimensional coordinate $E_r$ in the double-well potential $\Psi(E_r)$. 
A schematic structure of $\Psi$ is shown in Fig.~\ref{fig:bistable_potential}.
Among three roots, denoted by $E_1$, $E_2$, and $E_3$, respectively, $E_2$ is unstable because $(\partial^2\Psi/\partial E_r^2)_{E_r=E_2}<0$ and therefore will not be realized. 
The stable solution $E_1$ is negative and referred to as the ion-root, whereas $E_3$ is positive and referred to as the electron-root.

If $J_r$ depends on time only through $E_r$ and (\ref{eq:ambipolar_Er_evolution}) is given as a deterministic equation, the stationary solution is simply determined by the initial condition.
If $E_r<E_2$ initially, the system converges to $E_1$, whereas if $E_2<E_r$, it converges to $E_3$.
In the point particle picture, the representing point simply rolls down the potential hill to the nearest equilibrium point.
On the other hand, if a fluctuation with a sufficiently large amplitude is included in (\ref{eq:ambipolar_Er_evolution}), jumps between the two stable solutions over the potential barrier (the local maximum at $E_2$) are allowed. 

\cite{shaing1984stability} performed a stability analysis of the kind described above and predicted that, when fluctuations are considered, $E_r$ converges to the most stable root at $t\to\infty$.
Since the relative stability can be judged by the sign of $\Psi(E_3)-\Psi(E_1)$, this prediction is often represented as a root selection criterion in the form
\begin{equation}
\label{eq:Psi_E_critetrion}
\begin{split}
    \int_{E_1}^{E_3} J_r(E_r')& dE_r'
    =
    \Psi(E_3)-\Psi(E_1) \\
    &
\begin{cases}
    >0 & E_1 \ \text{(ion root) will be selected} \\
    =0 & \text{transition location} \\
    <0 & E_3 \ \text{(electron root) will be selected}
\end{cases}    
\end{split}
\end{equation}
In dynamical numerical simulations, the initial fluctuations and oscillations, including both numerical noise and physical processes such as geodesic acoustic mode (GAM) oscillation, effectively remove the initial condition dependency, while the fluctuation amplitudes decrease as the system approaches to a quasi-steady state.
Such a nature enables us to make quasi-deterministic predictions based on the criterion (\ref{eq:Psi_E_critetrion}).

From the definition, the integral (\ref{eq:Psi_E_critetrion}) corresponds to the area enclosed by the curve in the space of the coordinates $J_r$ and $E_r$, as illustrated, for example, in Figure.~1 of \citep{kuczynski2025assessment}.
Then, the determination of the transition position can be visualized by the equal-area rule analogous to the Maxwell construction for determining the phase equilibrium condition of a van der Waals gas \citep{landau1980statistical,atkins2023atkins} or similar Maxwell-type constructions for bistable chemical reaction systems \citep{schlogl1972chemical,keizer1978maxwell}.

%----------------
\subsection{Stochastic model}
%----------------

A more rigorous analysis on the effects of noise was performed by \cite{shaing1984noise}.
It should be noted first that, even without explicitly adding a noise term, the evolution equation (\ref{eq:ambipolar_Er_evolution}) is effectively stochastic. 
From a macroscopic perspective, the potential $\Psi$ is determined by stationary fluxes $\Gamma_a$, which depend on time only through $E_r$.
Microscopically, however, these fluxes are determined by the distribution function $f_a$, which involves stochasticity associated with thermal, numerical and/or turbulent fluctuations.
As a result, in real plasmas or Monte-Carlo kinetic simulations, stochasticity is inevitably present in (\ref{eq:ambipolar_Er_evolution}).

Thus, regardless of the presence of external fluctuation terms, the radial current may always be written as the sum of a deterministic part and a stochastic part,
\begin{equation}
\label{eq:current_decomposition}
    J_r 
    =
    \frac{\partial}{\partial E_r}\Psi
    +
    \tilde{J}_r,
\end{equation}
where the first term corresponds to stationary fluxes, while the stochastic term, $\tilde{J}_r$, may have a non-negligible contribution during the evolution. 
%Therefore, until the system approaches sufficiently close to a steady state, even when the background density and temperature profiles are fixed, the state of the system cannot be represented by a point moving along the curve of $\Psi$ depicted in Fig.~\ref{fig:bistable_potential}.

Assuming a Gaussian white noise, the equation becomes the overdamped Langevin equation
\begin{equation}
\label{eq:additive_langevin_E}
    \varepsilon\frac{\partial E_r}{\partial t}
    =
    -\frac{\partial\Psi}{\partial E_r}
    +\sqrt{2T_\text{eff}\varepsilon}\xi,
\end{equation}
where $T_\text{eff}$ is the noise amplitude, $\xi$ is a random variable that satisfies 
\begin{equation}
    \ev{\xi(t_1)\xi(t_2)}
    =\delta(t_1-t_2),
\end{equation}
and $\ev{...}$ denotes an ensemble average.
This equation describes the Brownian motion of an overdamped particle in the potential $\Psi$ if we interpret $\varepsilon$ as the friction coefficient and $T_\text{eff}$ as the (effective) temperature.

The corresponding Fokker-Planck equation for the probability density $p(E_r,t)$ is \citep{risken1984fokker}
\begin{equation}
    \frac{\partial}{\partial t}p(E_r,t)
    =
    \frac{\partial}{\partial E_r}
    \left[
    \frac{1}{\varepsilon}
    \Psi'(E_r) p(E_r,t)
    +
    D
    \frac{\partial p(E_r,t)}{\partial E_r}
    \right],
\end{equation}
where $D=T_\text{eff}/\varepsilon$ and $\Psi'=\partial\Psi/\partial E_r$.
The stationary solution has the form
\begin{equation}
\label{eq:additive_stationary_solution}
    p(E_r)
    =
    C\exp(-\frac{\Psi(E_r)}{T_\text{eff}}),
\end{equation}
where $C$ is the normalization coefficient.
Thus, $p(E_r)$ becomes a bimodal distribution corresponding to the bistability.
As the fluctuation amplitude decreases, the probability distribution shows a sharper peak around the most stable root $E_r^*$, and at the vanishing limit $T_\text{eff}\to 0$, it reduces to the delta function $\delta(E_r-E_r^*)$.

The problem of estimating the rate of transitions between stable states is known as the Kramers problem \citep{zwanzig2001nonequilibrium}.
When the noise is weak, as in quasi-steady states of neoclassical systems, the escape time from a stable state (or equivalently, mean first passage time to the local maximum of the potential) can be estimated by the well known Kramers' formula \citep{kramers1940brownian}[see also e.g., \citep{gardiner2009handbook,risken1984fokker} for derivations]
\begin{equation}
\label{eq:Kramers_escape_time}
    \tau_K
    =
    \frac{2\pi\varepsilon}{\sqrt{\omega_\text{min}|\omega_\text{max}|}}
    e^{\Delta \Psi/T_\text{eff}},
\end{equation}
where $\omega_\text{min}$ and $\omega_\text{max}$ are curvatures of the potential, $\omega\equiv \partial^2\Psi/\partial E_r^2$, at the position of minimum $E_\text{min}$ ($E_1$ or $E_3$ in Fig.~\ref{fig:bistable_potential}) and the local maximum $E_\text{max}$ ($E_2$ in Fig.~\ref{fig:bistable_potential}), respectively, and $\Delta \Psi=\Psi(E_\text{max})-\Psi(E_\text{min})$. 
This formula is valid under the assumption $T_\text{eff}\ll \Delta \Psi$.
If the escape time is sufficiently longer than the characteristic time scale of the neoclassical system, the root is stable and transition to another root is unlikely to be observed.
These stochastic descriptions are the essence of \cite{shaing1984noise}.

\subsection{Impact of the magnetic drift}

Now we discuss the main claim of this work. 
According to the definition of the potential function (\ref{eq:generalized_potential_Er}), the relative stability between $E_1$ and $E_3$ depends on the profile of the current $J_r$ in the interval $[E_1,E_3]$ crossing $E_r=0$.
In a conventional local simulation result, a steep potential barrier is formed due to a sharp peak of ion flux $\Gamma_i$ in this interval.
However, it has been shown that the ion flux in a nonaxisymmetric system strongly depends on the orbit model in the small-$E_r$ region.
Particularly, the inclusion of the magnetic drift prevents orbit resonance and suppresses the ion flux peaking in the regions significantly \citep{matsuoka2015effects,huang2017benchmark,velasco2020knosos,fujita2020global}.
Therefore, the size of the radial current will also be suppressed and the potential landscape is modified significantly. 
Depending on the situation, the choice of the orbit model may even change the selected root.
In the next section, we perform numerical simulations with different orbit models to demonstrate this effect.

Note that the noise amplitude of our neoclassical simulation at quasi-stationary states is sufficiently low, and we do not consider additional noise terms in the present study.
The transition behaviors will not be thus discussed for the following simulation results.
However, the modification of the potential landscape by the magnetic drift effect also changes the escape rate, and its magnitude may be significant.
We will discuss this at the end of this article.
\section{Confirmation by numerical simulations\label{sec:simulation}}

\subsection{Simulation models and setups}

For the analysis, we use the \textit{radially local} version of the neoclassical simulation code FORTEC-3D \citep{satake2006non,satake2008development}. 
While the original version of FORTEC-3D is a radially global code, the local version can solve several local orbit models including the zero-orbit-width (ZOW) model, in which the magnetic drift tangential to the flux surface is retained in the orbit equation \citep{matsuoka2015effects,huang2017benchmark}.
Using the ZOW model and a conventional local model neglecting the magnetic drift motion, we investigate how the magnetic drift modifies the potential structure governing the ambipolar electric field. 

The simulations are performed for a Large Helical Device (LHD) configuration with major radius $R_0=3.7 \ \mathrm{m}$, minor radius $a=0.63 \ \mathrm{m}$, and magnetic field strength at the magnetic axis $B_\text{ax}=2.37 \ \mathrm{T}$.
The plasma consists of electrons and hydrogen ions, and simulations are carried out on a flux surface at the normalized radial position $\rho=r/a=0.42$, where $r$ is the radial coordinate.
The profiles of the background density $n_0$ and temperature $T$, together with their radial gradients, are listed in Table.~\ref{tab:background_prof}.
The permittivity is $\varepsilon=4.6\times 10^{-10} \ \mathrm{F/m}$.
The basic collision time $\hat{\tau}_{ab}=4\pi \varepsilon_0 m_a^2v_{Ta}^3/(n_{b0}Z_a^2Z_b^2e^2\ln \Lambda_{ab})$ for each species combination is summarized in Table.~\ref{tab:basic_collision_time}, where $\varepsilon_0$ is the vacuum permittivity, $v_{Ta}=\sqrt{2T_a/m_a}$ is the thermal speed, and $\ln\Lambda_{ab}$ is the Coulomb logarithm.
FORTEC-3D is a Particle-In-Cell (PIC) code, and each simulation uses $1.28\times 10^6$ marker particles.

\begin{table}[h]
    \centering
    \begin{tabular}{c|cccc}
        Species & 
        $n_0 \ [\mathrm{1/m^3}]$ &
        $T_0 \ [\mathrm{keV}]$ &
        $d\ln n_0/dr \ [\mathrm{1/m}]$& 
        $d\ln T/dr \ [\mathrm{1/m}]$ \\ \hline
        Electron &
        $9.3\times 10^{18}$ &
        $4.1 $ &
        $-0.5$  &
        $-1.5$  \\
        Ion (hydrogen) &
        $9.3\times 10^{18}$ &
        $2.8$ &
        $-0.2$ &
        $-0.5$  \\
    \end{tabular}
    \caption{Profiles of the background density $n_0$ and $T$ of electrons and ions at the normalized radial position $\rho=0.42$ in the simulated LHD plasma.}
    \label{tab:background_prof}
\end{table}

\begin{table}[h]
    \centering
    \begin{tabular}{r|cc}
       \diagbox{a}{b}  & Electrons & Ions \\ \hline
       Electrons  & $4.2\times10^{-4}$ &  $4.1\times10^{-4}$ \\
       Ions       & $1.0\times10^{-2}$ & $8.9\times10^{-3}$ 
    \end{tabular}
    \caption{Basic collision time $\hat{\tau}_{ab}\ [\mathrm{s}]$ for each species combination.}
    \label{tab:basic_collision_time}
\end{table}

\subsection{Conventional local simulations}

\begin{figure}[h]
    \centering
    \includegraphics[width=0.6\textwidth,page=1]{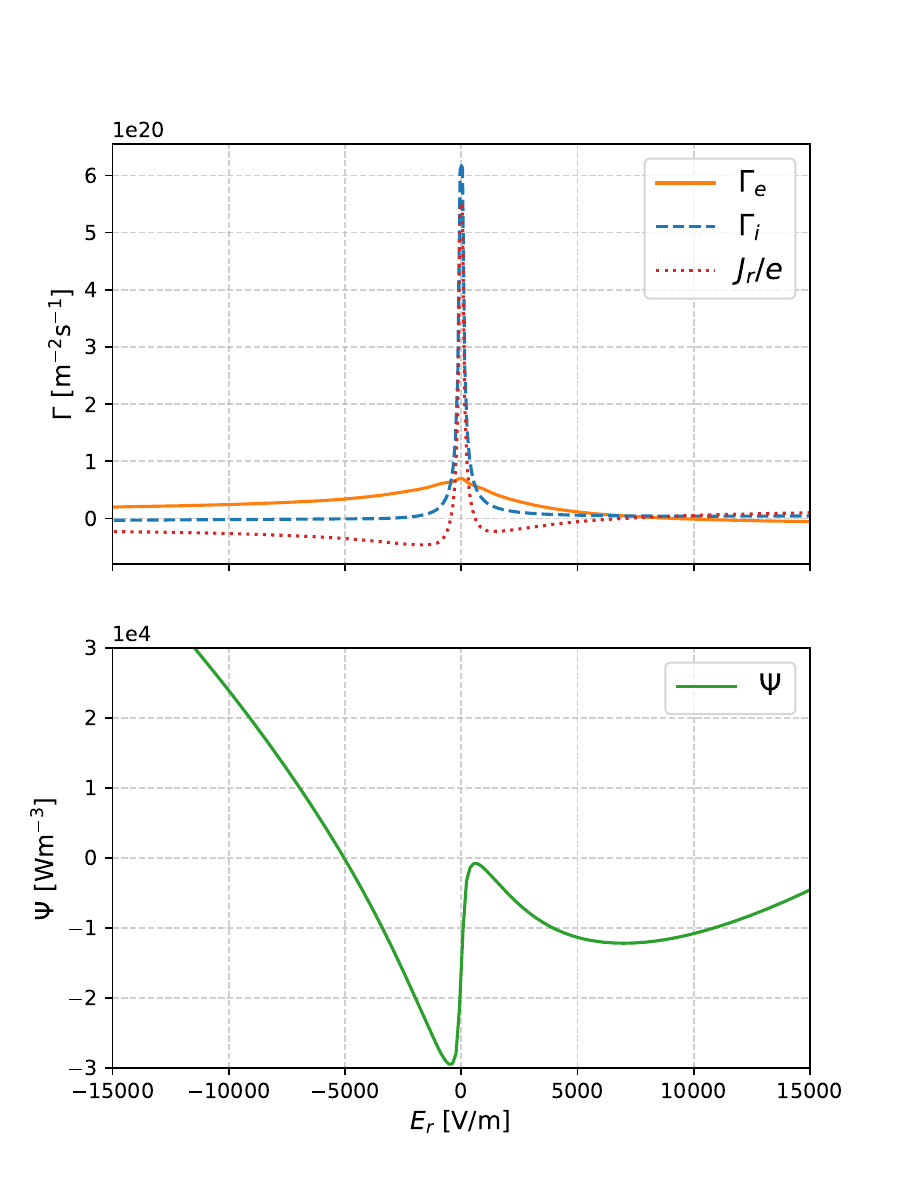}
    \caption{
    Upper panel: electron particle flux $\Gamma_e$ (orange solid line), ion particle flux $\Gamma_i$ (blue dashed line), and the radial current $J_r/e$ (red dotted line) as functions of the radial electric field $E_r$, obtained from DKES/PENTA calculations at $\rho=0.42$.
    Lower panel: potential function $\Psi(E_r)$ constructed from the radial current profile shown in the upper panel.
    The reference of $\Psi$ is chosen such that its local maximum is set to zero.
    }
    \label{fig:DKES_result_Erscan}
\end{figure}

Fig.~\ref{fig:DKES_result_Erscan} shows the $E_r$ dependence of the particle fluxes and radial current, as well as the corresponding potential landscape, obtained using the DKES/PENTA codes \citep{van1989variational,hirshman1986plasma}.
The potential stricture is constructed from the current profile using linear interpolation.
The reference level of the potential is chosen such that the local maximum is set to zero.
The DKES code computes the mono-energetic transport coefficients for given $E_r$, and PENTA constructs the neoclassical fluxes from the DKES results as functions of $E_r$.
In the upper figure, a sharp peak in the ion particle flux appears in the small-$E_r$ region, leading to the formation of a steep potential barrier in the corresponding region of the potential shown in the lower figure.
From the crossing points of $\Gamma_e$ and $\Gamma_i$ in the upper figure, or equivalently from the minima of the potential function, two stable roots are identified: an ion-root near $E_r\sim -450 \ \mathrm{V/m}$ and an electron-root near $E_r\sim 7 \ \mathrm{kV/m}$.
A comparison of potential well depths indicates that the ion-root is more stable than the electron root.
According to the criterion (\ref{eq:Psi_E_critetrion}), the ion-root is therefore expected to be selected.

\begin{figure}[h]
    \centering
    \includegraphics[width=0.6\textwidth,page=1]{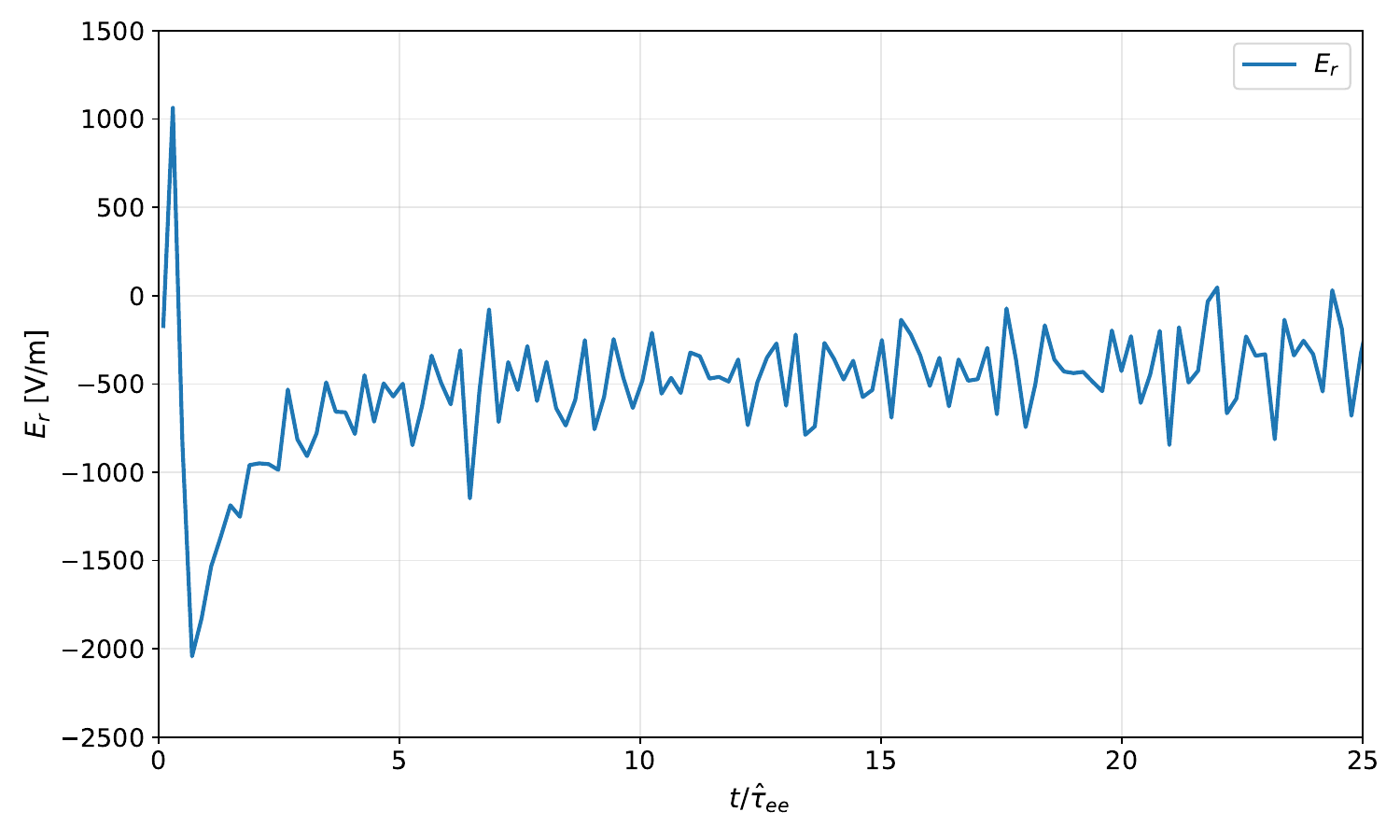}
    \caption{
    Time Evolution of the radial electric field $E_r$ obtained from a radially local FORTEC-3D simulation using the DKES-like model.
    The system evolves toward the ion-root solution predicted by the potential structure shown in Fig.~\ref{fig:DKES_result_Erscan}.}
    \label{fig:evol_Er_DKES}
\end{figure}

To confirm this prediction, we performed a local FORTEC-3D simulation in which conventional local drift-kinetic equations without the magnetic drift effect for both electrons and ions are solved self-consistently with the evolution for the ambipolar electric field.
The model used for the simulation is called the DKES-like model, and the details are found in \citep{matsuoka2015effects,huang2017benchmark}.
In the simulation, the ion-electron collisions are neglected, and for electron-ion collisions, the flux surface averaged ion parallel flow tabulated as a function of $E_r$ by PENTA is used. 
Starting with the initial condition $E_r=0$, the system evolves toward the ion-root state, as shown in Fig.~\ref{fig:evol_Er_DKES}. 
In the figure, the horizontal axis corresponds to time normalized by the basic electron collision time $\hat{\tau}_{ee}$.
The curve represents the short-time averaged value of $E_r$, where the average is taken over $\sim 0.2\hat{\tau}_{ee}$.
%The red curve represents the short-time averaged value of $E_r$, where the average is taken over $0.1\tau_e$.

\subsection{Simulations with the magnetic drift effect}

\begin{figure}[h]
    \centering
    \includegraphics[width=0.6\textwidth,page=1]{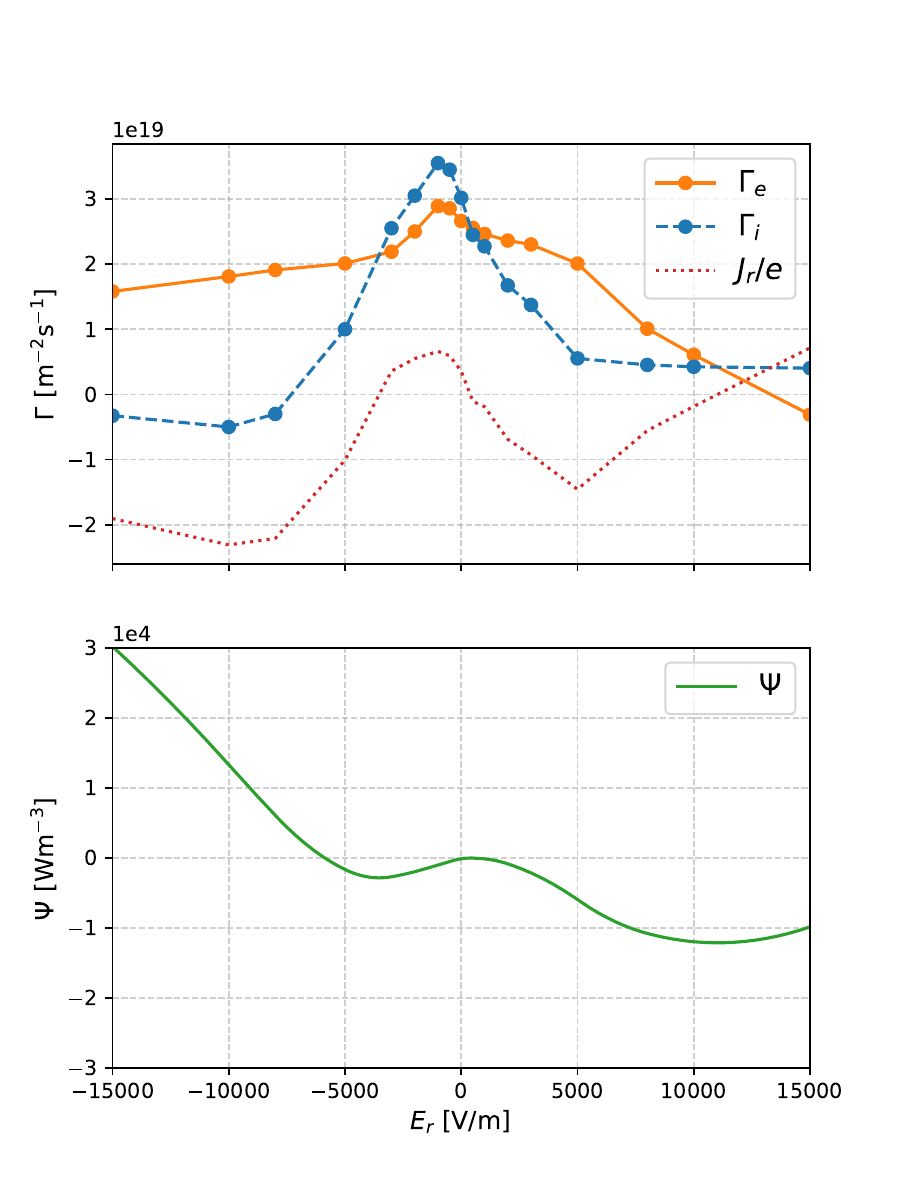}
    \caption{
    Upper panel: electron particle flux $\Gamma_e$ (orange solid line), ion particle flux $\Gamma_i$ (blue dashed line), and the radial current $J_r/e$ (red dotted line) as functions of the radial electric field $E_r$, obtained from radially local FORTEC-3D simulations using the zero-orbit-width (ZOW) model.
    Lower panel: potential function $\Psi(E_r)$ constructed from the radial current profile in the upper panel.
    The reference of $\Psi$ is chosen such that its local maximum is set to zero.}
    \label{fig:ZOW_result_Erscan}
\end{figure}

As a typical feature of conventional local simulations of low collisional non-axisymmetric plasmas, a strong ion flux peaking at $E_r\sim 0$ is found in the case discussed above. 
The deep potential well at the ion-root position relative to the electron-root position and the high potential barrier between the roots are both consequence of the overestimation of the ion flux. 
In such a situation, the inclusion of the magnetic drift is expected to have a qualitative impact on the relative stability of the roots. 
To examine the effects, we performed a series of local FORTEC-3D simulations using the ZOW model for fixed values of $E_r$.
Since the magnetic drift effect is small for the ion parallel flow \citep{huang2017benchmark}, the same ion flow data produced by PENTA as the conventional local simulation above is used for electron-ion collisions. 
The resulting fluxes and potential structure are shown in Fig.~\ref{fig:ZOW_result_Erscan}.
Compared with the DKES results, the ion flux peaking in the small-$E_r$ region is significantly suppressed, leading to a substantial modification of the potential structure.
Note that the electron flux is also reduced although less significant as the ion flux. 
As a consequence, the relative stability of the two roots is reversed, and the electron-root is the most stable state with the value $\sim 12 \ \mathrm{keV}$.
According to the root-selection criterion, the electron root is expected to be selected by this model. 

\begin{figure}[h]
    \centering
    \includegraphics[width=0.6\textwidth,page=1]{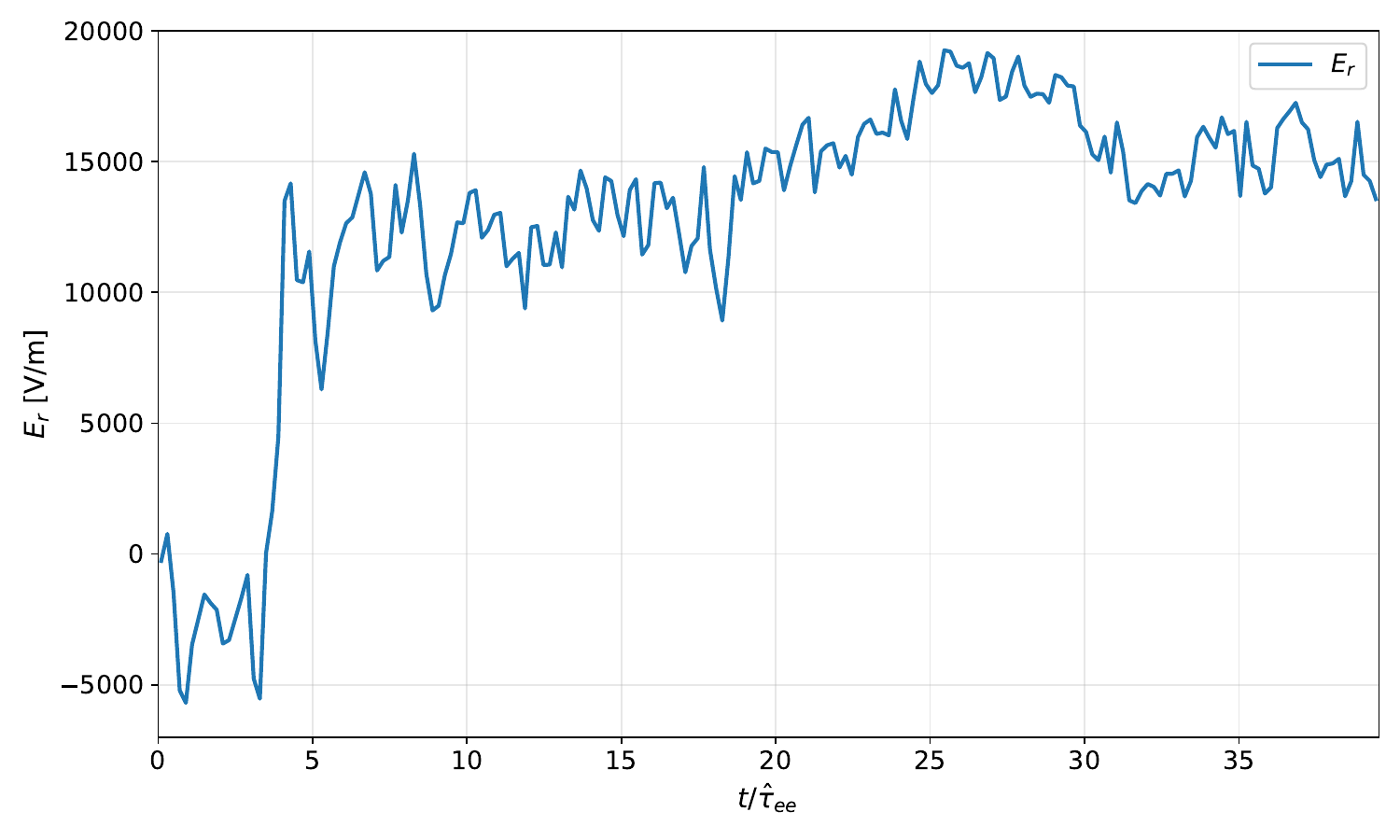}
    \caption{
    Time Evolution of the radial electric field $E_r$ obtained from a radially local FORTEC-3D simulation using the ZOW model.
    The system evolves toward the electron-root solution predicted by the potential structure shown in Fig.~\ref{fig:ZOW_result_Erscan}.}
    \label{fig:evol_Er_ZOW}
\end{figure}

To confirm the prediction, we solved the ZOW-type drift-kinetic equations for both species and the evolution of the ambipolar electric field using the local FORTEC-3D code.
Except for the orbit model, the simulation setups are the same as the local simulations using the DKES-like model above.
The time evolution of $E_r$ is shown in Fig.~\ref{fig:evol_Er_ZOW}, where the short-time average is taken over $0.2\hat{\tau}_{ee}$.
After the initial oscillation, the electric field evolves into the electron-root predicted by the potential structure.

These results clearly demonstrate that the inclusion of the magnetic drift can qualitatively alter the potential landscape and, consequently, the root selected by the neoclassical system.

%Note that changing from an electron-root to an ion-root is also possible. 
%For example, it has been shown that the inclusion of the magnetic drift lower the electron flux in large and positive $E_r$ region while keeping the ion flux almost unaffected \citep{huang2017benchmark}.
%In such cases, stable electron-root solutions may vanish, leaving only ion-root solutions.
\section{Summary and discussion\label{sec:summary}}

In this study, we discussed the impact of the magnetic drift in the orbit equation on the ambipolar condition.
For this purpose, we first reviewed the basic stability theory of thermodynamic systems. 
From this perspective, we clarified the role of the ambipolar condition in neoclassical transport theory: the ambipolar condition determines the steady state of the neoclassical system, and the stability of its root corresponds to the stability of the entire system.
These properties can be characterized by a simple one-dimensional potential function.

Revisiting the pioneering works by Shaing that clarified the structure of the ambipolar electric field dynamics, we explained the importance of the magnetic drift for correctly describing the dynamics of the ambipolar electric field.
In particular, inclusion of the magnetic drift may significantly modify the potential landscape and change the most stable root of the ambipolar condition.
To demonstrate this point, we performed numerical simulations and presented a case in which the most stable root changes from an ion-root to an electron-root as the magnetic drift modifies the potential structure.
This result provides a possible explanation for some discrepancies between different simulation models, such as the differences in root transition locations, and possibly between simulation results and observations.

Together with the underlying thermodynamic argument, these results highlight the importance of analyzing the nonaxisymmetric neoclassical systems in terms of the potential function $\Psi$.
This, in turn, emphasizes the necessity of including the magnetic drift effects when evaluating the potential structure, since such effect should be present in real plasmas.
The importance is therefore not limited to simulations exploring the small-$E_r$ region, but extends to the determination of steady states in nonaxisymmetric neoclassical systems in general.

This study also has further implications regarding the susceptibility of the ambipolar electric field to fluctuations.
If we apply an additive noise with the amplitude $T_\text{eff}=10^3 \ \mathrm{Wm^{-3}}$, the escape time for the DKES result is estimated using the Kramers formula (\ref{eq:Kramers_escape_time}) as
\begin{equation}
    \tau_\text{DKES}
    \sim 10^5 \ \mathrm{sec},
\end{equation}
which is far longer than the ion collision time $\sim 10^{-2} \ \mathrm{sec}$ and confinement time. 
On the other hand, for the same value of $T_\text{eff}$, the escape time for the ZOW result is estimated as
\begin{equation}
    \tau_\text{ZOW}
    \sim 10^{-1} \ \mathrm{sec},
\end{equation}
which is much closer to the collision time scale.
Moreover, global simulations further suppress the ion flux peaking \citep{huang2017benchmark}, which would further modify the potential structure. 
Fluctuations are prevalent in real plasmas and potentially relevant to the ambipolar electric field dynamics. 
For example, although the turbulent fluxes are ambipolar on average, they contribute to the evolution of the radial electric field as fluctuations, and they could even produce non-ambipolar components transiently \citep{idomura2021dynamics}.
These considerations suggest that the noise-induced transitions of the ambipolar electric field may be much more likely in real plasmas than previously expected.

An example indicating the role of fluctuations in radial electric field transition is presented by \cite{velasco2012vanishing,velasco2013damping}, where transitions in density-varying experiments are analyzed.
Although the paradigm of their problem is different from that in the present study, where the background density is fixed, the physical implications are closely related.
They showed that the near transition point in a density ramp-up discharge, the neoclassical viscosity (which corresponds to the potential curvature $\partial^2\Psi/\partial E_r^2$ in our study) vanishes, and the contribution of the fluctuating component of the radial current (which corresponds to the second term in (\ref{eq:current_decomposition})) becomes significant.
Generalized theoretical models to consider the variation of the background densities and temperatures for analyzing such problems will be presented in our future publications.

The susceptibility of the radial electric field to fluctuations also raises the possibility of controlling the ambipolar electric field profile through noise-induced transitions.
For example, if a transition from an ion-root to an electron root could be induced in the core region, even transiently, impurity ions could be efficiently exhausted from the core.
Thus, our findings have significant implications for device design and operation as well.
An additive white noise term has already been introduced to the evolution equation of $E_r$ in FORTEC-3D for investigating the GAM oscillation \citep{satake2011new}.
In future publications, we will generalize the form of the noise term and examine the impact of noise on the ambipolar electric field dynamics from more modern nonequilibrium thermodynamic perspectives and explore the practical feasibility of noise-induced control in more detail.

\begin{comment}
    From the Kramers formula, let us roughly estimate the escape time. 
$E\sim 10^3$, $J\sim 1$ thus $\omega\sim 10^{-3}$.
$eps sim 5*10^{-10}$.
Assuming $\Delta \Psi/T_\text{eff}\sim 10$ so $\exp(\Delta \Psi/T_\text{eff})\sim 10^4$, we have
\begin{equation}
    \tau_K
    \sim 
    10^{-1} 
\end{equation}
DKES:

omg~0.075,  eps~4.6*10^-10, exp[Delta Psi/T]~exp[30]~1*10^13
tau ~ 4*10^2 * 10^3 ~ 10^5

ZOW:
omg~0.001 eps~4.6*10^-10, exp[Delta Psi/T]~exp[10]~2.2*10^4
tau ~ 1000*2pi*4.6*10^{-10} *2*10^4
    ~ 6*10^4*10^{-10+4}~10^{-1}
\end{comment}

\section*{Acknowledgments}

We are grateful to Dr. J. L. Velasco for his valuable comments on the manuscript and for suggesting relevant references.

%\nocite{}
\bibliographystyle{apalike}
\bibliography{references}

%\end{multicols}

\end{document}